\documentstyle{article}
\newcommand{\C}{\mbox{\bf C}}
\newcommand{\R}{\mbox{\bf R}}
\newcommand{\M}{\mbox{M}}
\newcommand{\e}{\mbox{\bf e}}
\newcommand{\p}{\prime}
\newcommand{\s}{\scriptstyle}
\begin{document}
\begin{titlepage}
\title{{\huge Physical fields and Clifford algebras}}\vspace{1.2cm}
\author{Vadim V.Varlamov\\
{\small\it Applied Mathematics, Siberian State Academy of
Mining}\hspace{1mm}\small{\&}\hspace{1mm}{\small\it Metallurgy,}\\ 
{\small\it Novokuznetsk, Russia\thanks{E-mail: root@varlamov.kemerovo.su}}}
\date{}
\maketitle
\vspace{2cm}
\begin{abstract}The physical fields (electromagnetic and electron fields)
considered in the framework of Clifford algebras $\C_{2}$ and $\C_{4}$. The
electron field described by algebra $\C_{4}$ which in spinor representation
is realized by well-known Dirac $\gamma$-matrices, and by force of isomorphism
$\C_{4}\cong\C_{2}\otimes\C_{2}$ is represented as a tensor product of two
photon fields. By means of this introduced a system of electron field equations,
which in particular cases is coincide with Dirac's and Maxwell's equations.
\end{abstract}
\thispagestyle{empty}
\end{titlepage}
\newpage
%PACS numbers: 02.10.Tq, 03.50.De, 03.65.Pm

%\vspace{12mm}
It is well-known that Maxwell's equations
\begin{eqnarray}
\mbox{curl\bf E}+\dot{\mbox{\bf H}}&=&0,\nonumber \\
\mbox{div\bf H}&=&0,\nonumber \\
\mbox{curl\bf H}-\dot{\mbox{\bf E}}&=&\mbox{\bf j},\nonumber \\
\mbox{div\bf E}&=&\varrho\nonumber
\end{eqnarray}
and Dirac's equations
$$
(i\gamma_{\mu}\frac{\partial}{\partial x_{\mu}}-m)\psi=0
$$
may be rewriten in spinor form\cite{1}
$$
\begin{array}{ccc}
\partial^{\lambda\dot{\mu}}f_{\lambda}^{\rho}&=&0, \\
\partial^{\lambda\dot{\mu}}f_{\lambda}^{\rho}&=& s^{\rho\dot{\mu}}
\end{array}
$$
and
$$
\begin{array}{ccc}
\partial^{\lambda\dot{\mu}}\eta_{\dot{\mu}}+im\xi^{\lambda}&=&0, \\
\partial_{\lambda\dot{\mu}}\xi^{\lambda}+im\eta_{\dot{\mu}}&=&0,
\end{array}
$$
where
\begin{equation}\label{e4'}(\partial^{\lambda\dot{\mu}})=\left[
\begin{array}{cc}
\partial^{1\dot{1}} & \partial^{1\dot{2}} \\
\partial^{2\dot{1}} & \partial^{2\dot{2}}
\end{array}\right]=\left[
\begin{array}{cc}
\partial_{0}+\partial_{3} & \partial_{1}+i\partial_{2} \\
\partial_{1}-i\partial_{2} & \partial_{0}-\partial_{3}
\end{array}\right],
\end{equation}
\begin{equation}\label{e5}(f_{\lambda}^{\rho})=\left[
\begin{array}{cc}
f_{1}^{1} & f^{1}_{2} \\
f^{2}_{1} & f_{2}^{2}
\end{array}\right]=\left[
\begin{array}{cc}
F_{3} & F_{1}+iF_{2} \\
F_{1}-iF_{2} & -F_{3}
\end{array}\right]-\end{equation}
-the symmetric spin-tensor of complex electromagnetic field
$\mbox{\bf F}=\mbox{\bf E}+i\mbox{\bf H}$. The quantities $\xi^{\lambda}=
(\xi^{1},\xi^{2})$ and $\eta_{\dot{\mu}}=(\eta_{\dot{1}},\eta_{\dot{2}})$ are
make up a bispinor
$$\psi=\left[
\begin{array}{c}
\xi^{1} \\
\xi^{2} \\
\eta_{\dot{1}} \\
\eta_{\dot{2}}
\end{array}\right].$$

The spinors $\xi^{\lambda}$ and co-spinors $\eta_{\dot{\mu}}$ are satisfy to
the following conditions:
$$
\xi_{1}=\xi^{2},\hspace{2mm}\xi_{2}=-\xi^{1},\hspace{2mm}\eta^{\dot{1}}=
-\eta_{\dot{2}},\hspace{2mm}\eta^{\dot{2}}=\eta_{\dot{1}}.
$$

Besides, the quantities $\xi^{\lambda}=(\xi^{1},\xi^{2})$ and
$\eta_{\dot{\mu}}=(\eta_{\dot{1}},\eta_{\dot{2}})$ be vectors of two-dimensional
complex spaces (spin-spaces $S_{2}(i)$ and $\dot{S_{2}}(i)$); the each of
these spin-spaces is homeomorphic to extended complex plane. It is well-known
that the each spin-space be a space of linear representation of the some
Clifford algebra\cite{2}, in this case it is algebra $\C_{2}$ (so-called the algebra
of hyperbolic biquaternions). The motion group of each spin-spaces $S_{2}(i)$
and $\dot{S_{2}}(i)$ is isomorphic to a group $\mbox{SL}(2;\C)$ which be a
double-meaning representation of Lorentz group. By force of basic isomorphism
$\C_{2}\cong\M_{2}(\C)$ for the linear transformations of spinors of the
spaces $S_{2}(i)$ and $\dot{S_{2}}(i)$ we have:
$$\left[
\begin{array}{c}
{\xi^{1}}^{\prime} \\
{\xi^{2}}^{\prime}
\end{array}\right]=M\left[
\begin{array}{c}
\xi^{1} \\
\xi_{2}
\end{array}\right],
$$
$$\left[
\begin{array}{c}
{\eta_{\dot{1}}}^{\prime} \\
{\eta_{\dot{2}}}^{\prime}
\end{array}\right]=\dot{M}\left[
\begin{array}{c}
\eta_{\dot{1}} \\
\eta_{\dot{2}}
\end{array}\right],
$$
where $M,\dot{M}\in\M_{2}(\C)$.

Further on, consider a Clifford algebra $\R_{3}$ over a field of real numbers.
 The units of  this  algebra  are  satisfy  to  the  following
conditions:
$\e^{2}_{i}=1,\hspace{1mm}\e_{i}\e_{j}=-\e_{j}\e_{i}\hspace{2
mm}(i,j=1,2,3)$.

 Let
\begin{equation}\label{e9}\left.
\begin{array}{lcr}
{\cal A}_{0}&=&\partial_{0}\e_{0}+\partial_{1}\e_{1}+\partial_{2}\e_{2}+\partial_{3}\e_{3},\\
{\cal A}_{1}&=&A_{0}\e_{0}+A_{1}\e_{1}+A_{2}\e_{2}+A_{3}\e_{3},
\end{array}\right.
\end{equation}
where ${\cal A}_{0}$ and ${\cal A}_{1}$ be elements of $\R_{3}$. The coefficients
of these elements be partial derivatives and components of
vector-potential, respectively.

Make up now the exterior product of elements (\ref{e9}):
$${\cal A}_{0}{\cal A}_{1}=(\partial_{0}\e_{0}+\partial_{1}\e_{1}+\partial_{2}\e_{2}+\partial_{3}\e_{3})(A_{0}\e_{0}+A_{1}\e_{1}+A_{2}\e_{2}+A_{3}\e_{3})=$$
$$=(\underbrace{\partial_{0}A_{0}+\partial_{1}A_{1}+\partial_{2}A_{2}+\partial_{3}A_{3}}_{E_{0}})\e_{0}+(\underbrace{\partial_{0}A_{1}+\partial_{1}A_{0}}_{E_{1}})\e_{0}\e_{1}+$$
\begin{equation}\label{e10}
(\underbrace{\partial_{0}A_{2}+\partial_{2}A_{0}}_{E_{2}})\e_{0}\e_{2}+(\underbrace{\partial_{0}A_{3}+\partial_{3}A_{0}}_{E_{3}})\e_{0}\e_{3}+(\underbrace{\partial_{2}A_{3}-\partial_{3}A_{2}}_{H_{1}})\e_{2}\e_{3}+
\end{equation}
$$+(\underbrace{\partial_{3}A_{1}-\partial_{1}A_{3}}_{H_{2}})\e_{3}\e_{1}+(\underbrace{\partial_{1}A_{2}-\partial_{2}A_{1}}_{H_{3}})\e_{1}\e_{2}.$$

The scalar part $E_{0}\equiv 0$, since the first bracket in (\ref{e10}) be a Lorentz
condition $\partial_{0}A_{0}+\mbox{div}{\bf A}=0$. It is easily seen that the
other bracket be components of electric and magnetic fields:
$-E_{i}=-(\partial_{i}A_{0}+\partial_{0}A_{i}),\hspace{1mm}H_{i}=(\mbox{curl\bf A})_{i}$.

Since in this case the element $\omega=\e_{123}=\e_{1}\e_{2}\e_{3}$ is belong to a center of $\R_{3}$, then
\begin{equation}\label{e9'}
\omega \e_{1}=\e_{1}\omega=\e_{2}\e_{3},\hspace{2mm}\omega \e_{2}=\e_{2}\omega=\e_{3}\e_{1},
\hspace{2mm}\omega \e_{3}=\e_{3}\omega=\e_{1}\e_{2}.\end{equation}

In accordance with these correlations may be written (\ref{e10}) as
\begin{equation}\label{e11}
{\cal A}_{0}{\cal A}_{1}=F=(E_{1}+\omega H_{1})\e_{1}+(E_{2}+\omega H_{2})\e_{2}+(E_{3}+\omega H_{3})\e_{3}
\end{equation}

It is obvious that the expression (\ref{e11}) is coincide with the vector part
of complex quaternion (hyperbolic biquaternion) when $\e^{\p}_{1}=i\e_{1},
\e^{\p}_{2}=i\e_{2},\e^{\p}_{3}=i\e_{1}i\e_{2}$. Moreover, by general definition,
in the case of $n$ is odd the element $\omega=\e_{12\ldots n}$ is belong to a
center of $\R_{n}$, and when $n=4m^{\p}-1\hspace{2mm}(m^{\p}=1,2,\ldots)$ by
force of $\omega^{2}=-1$ there is the identity $\omega=i$, where $i$ is
imaginary unit. Hence it follows that
$$\R_{4m^{\p}-1}=\C_{4m^{\p}-2}.$$
In particular case, when $m^{\p}=1$ we obtain $\R_{3}=\C_{2}$. Thus, in
accordance with (\ref{e10}) and (\ref{e11}) we have the algebra $\C_{2}$ with
general element ${\cal A}=F_{0}\e_{0}+F_{1}\e_{1}+F_{2}\e_{2}+F_{3}\e_{3}$,
where $F_{0}=\partial_{0}A_{0}+\mbox{div\bf A}\equiv 0$ and $F_{i}\hspace{2mm}(i=1,2,3)-$ the
components of complex electromagnetic field.

Further on, make up the exterior product $\bigtriangledown\mbox{\bf F}$, where
$\bigtriangledown$ is the first element from (\ref{e9}) and {\bf F} is an expression
of type (\ref{e11}):
$$\bigtriangledown\mbox{\bf F}=\mbox{div\bf E}\e_{0}-((\mbox{curl\bf H})_{1}-\partial_{0}E_{1})\e_{1}-((\mbox{curl\bf H})_{2}-\partial_{0}E_{2})\e_{2}-$$
\begin{equation}\label{e12}
-((\mbox{curl\bf H})_{3}-\partial_{0}E_{3})\e_{3}+((\mbox{curl\bf E})_{1}+\partial_{0}H_{1})\e_{2}\e_{3}+((\mbox{curl\bf E})_{2}+\partial_{0}H_{2})\e_{3}\e_{1}+
\end{equation}
$$+((\mbox{curl\bf E})_{3}+\partial_{0}H_{3})\e_{1}\e_{2}+\mbox{div\bf H}\e_{1}\e_{2}\e_{3}.$$

It is easily seen that the first coefficient of the product $\bigtriangledown\mbox{\bf F}$
be a left part of equation $\mbox{div\bf E}=\varrho$. The following three
coefficients are make up a left part of equation $\mbox{curl\bf H}-\partial_{0}\mbox{\bf E}=j$,
the other coefficients are make up the equations $\mbox{curl\bf E}+\partial_{0}\mbox{\bf H}=0$
and $\mbox{div\bf H}=0$, respectively.

In accordance with (\ref{e9'}) this product may be rewritten as
$$\bigtriangledown\mbox{\bf F}=(\mbox{div\bf E}+\omega\mbox{div\bf H})\e_{0}+(-((\mbox{curl\bf H})_{1}-\partial_{0}E_{1})+\omega((\mbox{curl\bf E})_{1}+\partial_{0}H_{1}))\e_{1}+$$
$$+(-((\mbox{curl\bf H})_{2}-\partial_{0}E_{2})+\omega((\mbox{curl\bf E})_{2}+\partial_{0}H_{2})\e_{2}+$$$$+(-((\mbox{curl\bf H})_{3}-\partial_{0}E_{3})+\omega((\mbox{curl\bf E})_{3}+\partial_{0}H_{3}))\e_{3}.$$

It is obvious that in spinor representation of algebra $\R_{3}$ by force of
identity $\R_{3}=\C_{2}$ we have an isomorphism $\R_{3}\cong\M_{2}(\C)$.
At this isomorphism the units $\e_{i}\hspace{2mm}(i=0,1,2,3)$ of $\R_{3}$ are
correspond to the basis matrices of full matrix algebra $\M_{2}(\C)$:
\begin{equation}\label{e13}\left[
\begin{array}{cc}
1 & 0 \\
0 & 1
\end{array}\right],\hspace{2mm}\left[
\begin{array}{cc}
0 & 1 \\
1 & 0
\end{array}\right],\hspace{2mm}\left[
\begin{array}{cc}
0 & i \\
-i & 0
\end{array}\right],\hspace{2mm}\left[
\begin{array}{cc}
1 & 0 \\
0 & -1
\end{array}\right].\end{equation}
Hence it immediately follows that in spinor representation the first element
(\ref{e9}) in the base (\ref{e13}) has a form (\ref{e4'}), and, analoguosly,
for the vector-potential {\bf A} we have:
$$(a_{\lambda\dot{\mu}})=\left[
\begin{array}{cc}
a_{1\dot{1}} & a_{1\dot{2}} \\
a_{2\dot{1}} & a_{1\dot{2}}
\end{array}\right]=\left[
\begin{array}{cc}
A_{0}+A_{3} & A_{1}+iA_{2} \\
A_{1}-iA_{2} & A_{0}-A_{3}
\end{array}\right].$$
Thus, for the matrix of spin-tensor $f_{\lambda}^{\rho}$ and Maxwell's equations
we obtain the following expressions:
$$(f_{\lambda}^{\rho})=(\partial^{\rho\dot{\sigma}}a_{\lambda\dot{\sigma}})=\left[
\begin{array}{cc}
\partial^{1\dot{1}} & \partial^{1\dot{2}} \\
\partial^{2\dot{1}} & \partial^{2\dot{2}}
\end{array}\right]\left[
\begin{array}{cc}
a_{1\dot{1}} & a_{1\dot{2}} \\
a_{2\dot{1}} & a_{2\dot{2}}
\end{array}\right]=$$
$$\left[
\begin{array}{cc}
\partial_{0}+\partial_{3} & \partial_{1}+i\partial_{2} \\
\partial_{1}-i\partial_{2} & \partial_{0}-\partial_{3}
\end{array}\right]\left[
\begin{array}{cc}
A_{0}+A_{3} & A_{1}+iA_{2} \\
A_{1}-iA_{2} & A_{0}-A_{3}
\end{array}\right]=\left[
\begin{array}{cc}
F_{3} & F_{1}+iF_{2} \\
F_{1}-iF_{2} & -F_{3}
\end{array}\right],$$
$$(s^{\rho\dot{\mu}})=(\partial^{\lambda\dot{\mu}}f_{\lambda}^{\rho})=\left[
\begin{array}{cc}
\partial_{0}+\partial_{3} & \partial_{1}+i\partial_{2} \\
\partial_{1}-i\partial_{2} & \partial_{0}-\partial_{3}
\end{array}\right]\left[
\begin{array}{cc}
F_{3} & F_{1}+iF_{2} \\
F_{1}-iF_{2} & -F_{3}
\end{array}\right]=$$
{\renewcommand{\arraystretch}{1.5}
$$\left[
\begin{tabular}{c|c}
$\s\mbox{\scriptsize div\bf E}-(\mbox{\scriptsize curl\bf H})_{3}+\partial_{0}E_{3}+$                             & $\s-(\mbox{\scriptsize curl\bf H})_{1}+\partial_{0}E_{1}+(\mbox{\scriptsize curl\bf E})_{2}+\partial_{0}H_{2}+$ \\
$\s+i(\mbox{\scriptsize div\bf H}+(\mbox{\scriptsize curl\bf E})_{3}+\partial_{0}H_{3})$                          & $\s+i((\mbox{\scriptsize curl\bf E})_{1}+\partial_{0}H_{1}-(\mbox{\scriptsize curl\bf H})_{2}+\partial_{0}E_{2})$ \\ \hline
$\s-(\mbox{\scriptsize curl\bf H})_{1}+\partial_{0}E_{1}+(\mbox{\scriptsize curl\bf E})_{2}+\partial_{0}H_{2}-$   & $\s\mbox{\scriptsize div\bf E}+(\mbox{\scriptsize curl\bf H})_{3}-\partial_{0}E_{3}-$                             \\
$\s-i((\mbox{\scriptsize curl\bf E})_{1}+\partial_{0}H_{1}-(\mbox{\scriptsize curl\bf H})_{2}+\partial_{0}E_{2})$ & $\s-i((\mbox{\scriptsize curl\bf E})_{3}+\partial_{0}H_{3}+\mbox{\scriptsize div\bf H})$                            \\
\end{tabular}\right].$$}
A dual exterior product we obtain by force of identity $\ast=\omega$, where
$\ast$ is operator of Hodge\cite{3} and $\omega=\e_{12\ldots n}$ is
volume element of Clifford algebra $\R_{n}$. This identity is possible only if
$n=4m^{\p}+1$ or $n=4m^{\p}-1$, where $m^{\p}=~1,2,\ldots$; since only
in this case $\omega$ is belong to a center of $\R_{n}$. For $\R_{3}$ we have $\omega = \e_{123}$.
This way
\begin{eqnarray}
\ast({\cal A}_{0}{\cal A}_{1})&=&\omega({\cal A}_{0}{\cal A}_{1})=-H^{1}\e_{0}\e_{1}-H^{2}\e_{0}\e_{2}-H^{3}\e_{0}\e_{3}+\label{e14}\\
& & +E^{1}\e_{2}\e_{3}+E^{2}\e_{3}\e_{1}+E^{3}\e_{1}\e_{2}.\nonumber
\end{eqnarray}
In the base (\ref{e13}) for a complex conjugate electromagnetic field from
$\ast(i{\cal A}_{0}{\cal A}_{1})$ we have:
$$(f_{\dot{\lambda}}^{\dot{\rho}})=\left[
\begin{array}{cc}
f^{\dot{1}}_{\dot{1}} & f^{\dot{1}}_{\dot{2}} \\
f^{\dot{2}}_{\dot{1}} & f^{\dot{2}}_{\dot{2}}
\end{array}\right]=\left[
\begin{array}{cc}
\stackrel{\ast}{F_{3}} & \stackrel{\ast}{F_{1}}-i\hspace{-1mm}\stackrel{\ast}{F_{2}} \\
\stackrel{\ast}{F_{1}}+i\hspace{-1mm}\stackrel{\ast}{F_{2}} & -\stackrel{\ast}{F_{3}}
\end{array}\right],$$
where \raisebox{0.22ex}{$\stackrel{\ast}{{\bf F}}$}=${\bf E}-i{\bf H}$.

Accordingly, a system of complex conjugate Maxwell's equations may be written as
\begin{eqnarray}
\partial^{\mu\dot{\lambda}}f_{\dot{\lambda}}^{\dot{\rho}}&=&0,\nonumber \\
\partial^{\mu\dot{\lambda}}f_{\dot{\lambda}}^{\dot{\rho}}&=&s^{\mu\dot{\rho}}. \nonumber
\end{eqnarray}
In the terms of $\R_{3}=\C_{2}$ this system is equivalent to a system of
coefficients of exterior product $\bigtriangledown\hspace{-1mm}\stackrel{\ast}{F}$, where
$\stackrel{\ast}{F}$ is a dual exterior product of type (\ref{e14}).

It is easily verified that a stress-energy tensor of electromagnetic field
in spinor form is realized by spin-tensor $t_{\lambda\dot{\nu}}^{\rho\dot{\mu}}=
f_{\lambda}^{\rho}f_{\dot{\nu}}^{\dot{\mu}}$, the matrix of which has a form:
$$(t_{\lambda\dot{\nu}}^{\rho\dot{\mu}})=(f_{\lambda}^{\rho}f_{\dot{\nu}}^{\dot{\mu}})=
{\renewcommand{\arraystretch}{1.2}\left[
\begin{array}{cc}
F_{3} & F_{1}+iF_{2} \\
F_{1}-iF_{2} & -F_{3} 
\end{array}\right]}\left[
\begin{array}{cc}
\stackrel{\ast}{F_{3}} & \stackrel{\ast}{F_{1}}-i\hspace{-1mm}\stackrel{\ast}{F_{2}} \\
\stackrel{\ast}{F_{1}}+i\hspace{-1mm}\stackrel{\ast}{F_{2}} & -\stackrel{\ast}{F_{3}}
\end{array}\right]=$$
$${\renewcommand{\arraystretch}{1.7}\left[
\begin{tabular}{c|c}
$\s W-2s_{3}+i(\sigma_{21}-\sigma_{12})$ & $\s\sigma_{31}-2s_{1}-\sigma_{13}+i(\sigma_{32}-2s_{2}-\sigma_{23})$ \\ \hline
$\s\sigma_{13}-2s_{1}-\sigma_{31}+i(\sigma_{32}+2s_{2}-\sigma_{23})$ & $\s W+2s_{3}+i(\sigma_{12}-\sigma_{21})$ \\
\end{tabular}\right]},$$
where $W$ is density of energy, $\sigma_{ik}$ is stress tensor, and {\bf s} is the vector of Poynting.

Thus, {\it we have a full description of the basic notions of electromagnetic field
in the terms of algebra $\R_{3}=\C_{2}$ and its spinor representation.}

Consider now an algebra $\C_{4}$. In the spinor representation this algebra
is isomorphic to a matrix algebra $\M_{4}(\C)$, the base of which consist of
well-known Dirac $\gamma$-matrices. In the base of Weyl for these matrices we
have
$$\gamma^{m}=\left[
\begin{array}{cc}
0 & \sigma^{m} \\
\overline{\sigma}^{m} & 0
\end{array}\right],$$
where $m=0,1,2,3$ and
$$\sigma^{0}=\left[
\begin{array}{cc}
1 & 0 \\
0 & 1
\end{array}\right],\hspace{2mm}\sigma^{1}=\left[
\begin{array}{cc}
0 & 1 \\
1 & 0
\end{array}\right],\hspace{2mm}\sigma^{2}=\left[
\begin{array}{cc}
0 & -i \\
i & 0
\end{array}\right],\hspace{2mm}\sigma^{3}=\left[
\begin{array}{cc}
1 & 0 \\
0 & -1
\end{array}\right],$$
$$\overline{\sigma}^{0}=\sigma^{0},\hspace{2mm}\overline{\sigma}^{1,2,3}=-\sigma^{1,2,3}.$$
In this base the system of Dirac's equations has a following form:
$$\left[
\begin{array}{cccc}
0 & 0 & \partial^{1\dot{1}} & \partial^{2\dot{1}} \\
0 & 0 & \partial^{1\dot{2}} & \partial^{2\dot{2}} \\
\partial_{1\dot{1}} & \partial_{1\dot{2}} & 0 & 0 \\
\partial_{2\dot{1}} & \partial_{2\dot{2}} & 0 & 0
\end{array}\right]\left[
\begin{array}{c}
\xi^{1} \\
\xi^{2} \\
\eta_{\dot{1}} \\
\eta_{\dot{2}}
\end{array}\right]=-im\left[
\begin{array}{c}
\xi^{1} \\
\xi^{2} \\
\eta_{\dot{1}} \\
\eta_{\dot{2}}
\end{array}\right].$$
Replace the symmetric spin-tensors $\partial^{\mu\dot{\lambda}}$ and
$\partial_{\lambda\dot{\mu}}$ by the symmetric spin-tensors $s^{\mu\dot{\lambda}}$
and $s_{\lambda\dot{\mu}}$. Then
$$\left[
\begin{tabular}{c|c}
0  & $s^{\mu\dot{\lambda}}$  \\ \hline
$s_{\lambda\dot{\mu}}$ & 0 
\end{tabular}\right]\left[
\begin{array}{c}
\xi \\
\dot{\eta}
\end{array}\right]=$$
$$\left[{\renewcommand{\arraystretch}{1.3}
\begin{array}{cccc}
\s 0 & \s 0 & \s\partial^{1\dot{1}}f^{\dot{1}}_{\dot{1}}+\partial^{2\dot{1}}f^{\dot{2}}_{\dot{1}} & \s\partial^{1\dot{1}}f^{\dot{1}}_{\dot{2}}+\partial^{2\dot{1}}f^{\dot{2}}_{\dot{2}} \\
\s 0 & \s 0 & \s\partial^{1\dot{2}}f^{\dot{1}}_{\dot{1}}+\partial^{2\dot{2}}f^{\dot{2}}_{\dot{1}} & \s\partial^{1\dot{2}}f^{\dot{1}}_{\dot{2}}+\partial^{2\dot{2}}f^{\dot{2}}_{\dot{2}} \\
\s\partial_{1\dot{1}}f^{1}_{1}+\partial_{1\dot{2}}f^{2}_{1} & \s\partial_{1\dot{1}}f^{1}_{2}+\partial_{1\dot{2}}f^{2}_{2} &\s 0 &\s 0 \\
\s\partial_{2\dot{1}}f^{1}_{1}+\partial_{2\dot{2}}f^{2}_{1} & \s\partial_{2\dot{1}}f^{1}_{2}+\partial_{2\dot{2}}f^{2}_{2} &\s 0 &\s 0
\end{array}}\right]\hspace{-0.5mm}\left[
{\renewcommand{\arraystretch}{1.3}
\begin{array}{c}
\s\xi^{1} \\
\s\xi^{2} \\
\s\eta_{\dot{1}} \\
\s\eta_{\dot{2}} 
\end{array}}\right]=\s-im\left[
{\renewcommand{\arraystretch}{1.3}
\begin{array}{c}
\s\xi^{1} \\
\s\xi^{2} \\
\s\eta_{\dot{1}} \\
\s\eta_{\dot{2}}
\end{array}}\right].$$
Hence we obtain the following system of equations:
\begin{equation}\label{e15}
{\renewcommand{\arraystretch}{1.4}
\begin{array}{ccc}
(\partial^{1\dot{1}}f^{\dot{1}}_{\dot{1}}+\partial^{2\dot{1}}f^{\dot{2}}_{\dot{1}})\eta_{\dot{1}}+(\partial^{1\dot{1}}f^{\dot{1}}_{\dot{2}}+\partial^{2\dot{1}}f^{\dot{2}}_{\dot{2}})\eta_{\dot{2}} & = & -im\xi^{1}, \\
(\partial^{1\dot{2}}f^{\dot{1}}_{\dot{1}}+\partial^{2\dot{2}}f^{\dot{2}}_{\dot{1}})\eta_{\dot{1}}+(\partial^{1\dot{2}}f^{\dot{1}}_{\dot{2}}+\partial^{2\dot{2}}f^{\dot{2}}_{\dot{2}})\eta_{\dot{2}} & = & -im\xi^{2}, \\
(\partial_{1\dot{1}}f^{1}_{1}+\partial_{1\dot{2}}f^{2}_{1})\xi^{1}+(\partial_{1\dot{1}}f^{1}_{2}+\partial_{1\dot{2}}f^{2}_{2})\xi^{2} & = & -im\eta_{\dot{1}}, \\
(\partial_{2\dot{1}}f^{1}_{1}+\partial_{2\dot{2}}f^{2}_{1})\xi^{1}+(\partial_{2\dot{1}}f^{1}_{2}+\partial_{2\dot{2}}f^{2}_{2})\xi^{2} & = & -im\eta_{\dot{2}}.
\end{array}}\end{equation}

In particular case, when $f^{1}_{1}=f^{2}_{2}=f^{\dot{1}}_{\dot{1}}=
f^{\dot{2}}_{\dot{2}}=1$ and $f^{2}_{1}=f^{1}_{2}=f^{\dot{2}}_{\dot{1}}=
f^{\dot{1}}_{\dot{2}}=0$ system (\ref{e15}) is coincide with Dirac's equations
\begin{eqnarray}
\partial^{1\dot{1}}\eta_{\dot{1}}+\partial^{2\dot{1}}\eta_{\dot{2}}&=&-im\xi^{1},\nonumber \\
\partial^{1\dot{2}}\eta_{\dot{1}}+\partial^{2\dot{2}}\eta_{\dot{2}}&=&-im\xi^{2},\nonumber \\
\partial_{1\dot{1}}\xi^{1}+\partial_{1\dot{2}}\xi^{2}&=&-im\eta_{\dot{1}}, \nonumber \\
\partial_{2\dot{1}}\xi^{1}+\partial_{2\dot{2}}\xi^{2}&=&-im\eta_{\dot{2}}. \nonumber
\end{eqnarray}

Analoguosly, when $\eta_{\dot{1}}=\xi^{2}=1,\hspace{1mm}\eta_{\dot{2}}=\xi^{1}=0$ and
$m=0$ from (\ref{e15}) we obtain the following system of equations
\begin{eqnarray}
\partial^{1\dot{1}}f^{\dot{1}}_{\dot{1}}+\partial^{2\dot{1}}f^{\dot{2}}_{\dot{1}}&=&0, \nonumber \\
\partial^{1\dot{2}}f^{\dot{1}}_{\dot{1}}+\partial^{2\dot{2}}f^{\dot{2}}_{\dot{1}}&=&0, \nonumber \\
\partial_{1\dot{1}}f^{1}_{2}+\partial_{1\dot{2}}f^{2}_{2}&=&0, \nonumber \\
\partial_{2\dot{1}}f^{1}_{2}+\partial_{2\dot{2}}f^{2}_{2}&=&0. \nonumber
\end{eqnarray}
or
\begin{equation}\label{e18}
{\renewcommand{\arraystretch}{1.4}
\begin{array}{ccc}
(\partial_{0}+\partial_{3})\hspace{-1mm}\stackrel{\ast}{F_{3}}+(\partial_{1}-i\partial_{2})(\stackrel{\ast}{F_{1}}+i\hspace{-1mm}\stackrel{\ast}{F_{2}}) &=&0,  \\
(\partial_{1}+i\partial_{2})\hspace{-1mm}\stackrel{\ast}{F_{3}}+(\partial_{0}-\partial_{3})(\stackrel{\ast}{F_{1}}+i\hspace{-1mm}\stackrel{\ast}{F_{2}}) &=&0,  \\
(\partial_{0}+\partial_{3})(F_{1}+iF_{2})-(\partial_{1}+i\partial_{2})F_{3}&=& 0, \\
(\partial_{1}-i\partial_{2})(F_{1}+iF_{2})-(\partial_{0}-\partial_{3})F_{3}&=& 0. 
\end{array}}\end{equation}
Substitute in (\ref{e18}) $\stackrel{\ast}{\mbox{\bf F}}=\mbox{\bf E}
-i\mbox{\bf H}$, $\mbox{\bf F}=\mbox{\bf E}+i\mbox{\bf H}$ and divide the
real and imaginary parts after simple transformations we have
\begin{eqnarray}
& &\mbox{div\bf E}-(\mbox{curl\bf H})_{3}+\partial_{0}E_{3}=0, \nonumber \\
&-\hspace{-2mm}&\mbox{div\bf H}-(\mbox{curl\bf E})_{3}-\partial_{0}H_{3}=0, \nonumber  \\
&-\hspace{-2mm}&(\mbox{curl\bf H})_{1}+\partial_{0}E_{1}+(\mbox{curl\bf E})_{2}+\partial_{0}H_{2}=0, \nonumber \\
&-\hspace{-2mm}&(\mbox{curl\bf H})_{2}+\partial_{0}E_{2}-(\mbox{curl\bf E})_{1}-\partial_{0}H_{1}=0, \nonumber \\
&-\hspace{-2mm}&(\mbox{curl\bf H})_{1}+\partial_{0}E_{1}-(\mbox{curl\bf E})_{2}-\partial_{0}H_{2}=0, \nonumber \\
&-\hspace{-2mm}&(\mbox{curl\bf H})_{2}+\partial_{0}E_{2}+(\mbox{curl\bf E})_{1}+\partial_{0}H_{1}=0, \nonumber \\
& &\mbox{div\bf E}+(\mbox{curl\bf H})_{3}-\partial_{0}E_{3}=0, \nonumber \\
& &\mbox{div\bf H}-(\mbox{curl\bf E})_{3}-\partial_{0}H_{3}=0. \nonumber
\end{eqnarray}
From the latest equations by means of addition and substraction we obtain the
system of Maxwell's equations for empty space:
\begin{eqnarray}
&&\mbox{div\bf E}=0, \nonumber \\
&&\mbox{div\bf H}=0, \nonumber \\
&&\mbox{curl\bf H}-\partial_{0}\mbox{\bf E}=0, \nonumber \\
&&\mbox{curl\bf E}+\partial_{0}\mbox{\bf H}=0. \nonumber
\end{eqnarray}
It is obvious that we obtain the same result if suppose in system (\ref{e15})
$m=0$ and $\eta_{\dot{2}}=\xi^{1}=1,\hspace{1mm}\eta_{\dot{1}}=\xi^{2}=0$.

Thus, {\it system} (\ref{e15}) {\it which we shall call the system of electron field equations,
in the particular cases is coincide with Dirac's and Maxwell's equations}.

On the other hand, by force of isomorphism\cite{4} $\C_{4}\cong\C_{2}\otimes\C_{2}$
or $\C_{4}\cong\C_{2}\otimes\stackrel{\hspace{-1mm}\ast}{\C_{2}}$, where $\stackrel{\hspace{-1mm}\ast}{\C_{2}}$
is an algebra with general element \raisebox{0.3ex}{$\stackrel{\hspace{1mm}\ast}{{\cal A}}$}=
$\stackrel{\ast}{F_{0}}\e_{0}+\stackrel{\ast}{F_{1}}\e_{1}+\stackrel{\ast}{F_{2}}\e_{2}
+\stackrel{\ast}{F_{3}}\e_{12}$, we may say that {\it the electron field be a
tensor product of two photon fields.} Moreover, the algebras $\C_{2}$ and
$\stackrel{\hspace{-1mm}\ast}{\C_{2}}$ are represent the photon fields with 
left-handed and right-handed polarization, respectively (see \cite{5}).

\end{document}